\documentclass[final,longbibliography,12pt]{revtex4-1}
\usepackage{amsmath}  
\usepackage{amsfonts} 
\usepackage{graphicx} 
\usepackage{slashed} 

\newcommand{\be}{\begin{equation}}
\newcommand{\ee}{\end{equation}}
\newcommand{\wf}{wavefunction\;}
\newcommand{\wfs}{wavefunctions\;}
\newcommand{\infl}{influence function\;}
\newcommand{\infls}{influence functions\;}

\newcommand{\rmi}{\mathrm{i}}

\newcommand{\dg}{\dagger}
\newcommand{\fy}{\slashed}
\newcommand{\zo}{\mathbf 0}
\newcommand{\ve}{\mathbf e}

\begin{document}

\title{Duffin--Kemmer--Petiau particles {\it are}  bosons}

\author{A. F. Bennett}
\email{bennetan@oregonstate.edu} 
\affiliation{College of Earth, Ocean and Atmospheric Sciences\\
Oregon State University\\104 CEOAS Administration Building\\ Corvallis, OR 97331-5503, USA} 

\date{\today}

\author{A. F. Bennett}

\date{\today}

\begin{abstract}
The parametrized Duffin--Kemmer--Petiau wave equation is formulated here for many relativistic particles of spin--0 or spin--1. The conventional second-quantized or Fock-space proof of the spin-statistics connection requires that the fields of creation and annihilation operators satisfy commutation relations subject to causality conditions. The conditions restrict entanglement to spacelike separations of the order of the Compton wavelength $\hbar/mc$. Relativistic quantum mechanics is used here to prove the symmetry of the \wfs for identical particles, following the  nonrelativistic argument of Jabs (Found. Phys. 2010, {\bf 40}, 776--792).  First quantization does not require causal commutation relations, and so entanglement is unrestricted.

%
%
%

\end{abstract}

\maketitle

\section{Introduction}\label{S:intro}
The relativistic wave equation of Duffin \cite{Duff38}, Kemmer \cite{Kemm39} and Petiau \cite{Pet36} provides  a five--dimensional representation of spin--0 and a ten-dimensional representation of spin--1. The only fundamental particle of spin--0 that has been observed is the Higgs boson \cite{Cho12,Cho12E}. There are of course mesons of spin--0 and spin--1 \cite{Griff08}. Cooper pairs are typically in the singlet (spin--0) state, but pairs have  been observed in the triplet  (spin--1) state  \cite{Sprung2010}. The objective here is the establishment of Bose-Einstein statistics for  Duffin--Kemmer--Petiau (hereafter DKP) particles. 
The spin--statistics connection has of course long been established in Quantum Field Theory \cite{IZ,Wei95,Zee03,Sr07}, but the proof according to  Quantum Field Theory (hereafter QFT) requires that the creation and annihilation operators for particles of integer (half--odd--integer) spin satisfy commutation (anticommutation) relations subject to causality conditions.  As a consequence of these causal relations, quantum entanglement is maximal in the vacuum state of QFT and limited to spacetime separations of the order of the Compton wavelength $\hbar/mc$ \cite{Haag58,Araki62,Fred85,Summ11,Ols12}\,. The significance of the result given here is that the first-quantized formulation does not involve operator fields, there is  therefore no requirement for causal commutation relations,  and as a consequence  quantum entanglement is unrestricted. The formulation considered here  is not an acausal theory, however, as the range of influence for the parametrized DKP wave equation is shown below to be restricted to the surface and interior of the light cone.  

The relativistic proof of Bose--Einstein statistics given here extends the nonrelativistic proof given by Jabs \cite{Jabs10,Jabs2014}.  The role of coordinate time in Jab's proof is taken here by a real parameter, following the proof of Fermi--Dirac statistics for relativistic particles of spin--1/2 using the parametrized Dirac wave equation \cite{Alv98,John69,Fan93,Benn2014,Benn2014arX,Benn2015}.

Jabs' proof \cite{Jabs10,Jabs2014} has two aspects.  First, specifying a spin or rotation axis does not specify a frame. Thus two eigenstates with respect to the same spin operator or the same rotation operator may not be referred to the same frame. Homotopically consistent rotation to a common frame yields complex phase factors which imply that  \wfs for identical particles of  integer spin must be symmetric. The second aspect  in Jabs' proof  is the implicit understanding that rotation commutes with the nonrelativistic free Hamiltonian and so `once an eigenstate, always an eigenstate'.  Such is not the case even for free solutions of the  DKP wave equation. However, the eigenstates of rotation constitute a basis, and so any  solution  of  the DKP equation may be considered to have been prepared as a rotational eigenstate on some spacetime manifold at a particular value of the parameter $\tau$, say $\tau=0$\,. The first element in Jab's proof then establishes the symmetry of the DKP \wf  at $\tau=0$, and it is shown here that the  evolution of the \wf for $\tau >0$ (or for $\tau<0)$  preserves the  symmetry. The case of spin--1 is treated in full and  the case of spin--0 is closely analogous. The case of  spin--1/2 is also closely analogous, leading to Fermi--Dirac statistics in a simpler and more general way than in \cite{Benn2015}.

The outline of this article is as follows. The DKP equation is developed in \S\,\ref{S:DKP} from  Proca's equations for a massive classical vector field. The  `meson algebra' of the DKP coefficient matrices is stated and the covariance of the formulation is outlined. Plane \wfs and   \infls for the single--particle and two--particle parametrized DKP equations are given in \S\,\ref{S:pDKP}. The rotation operator for DKP spinors is introduced in \S\,\ref{S:prep}, together with its eigenstates. The latter provide a basis for all 10--spinors, and so it maybe assumed that a particle is in an eigenstate at $\tau=0$\,. Jabs' argument \cite{Jabs10,Jabs2014} for the symmetry of the prepared  \wf  for identical particles applies immediately, and the details are given here for two particles. Owing to the tensor structure of the parametrized DKP equation for many particles, the symmetry of the prepared \wf is preserved in the subsequent evolution for $\tau>0$\,. The entanglement implied by the preserved symmetry is discussed in \S\,\ref{S:ent}.    The proof of Bose--Einstein statistics for DKP particles is briefly compared in \S\,\ref{S:disc}  to the proof for integer--spin fields according to Quantum Field Theory.  Some details for spin--0 DKP particles and spin--1/2 Dirac particles are included in an appendix.

\section{The Duffin-Kemmer-Petiau wave equation}\label{S:DKP}

\subsection{Proca's equations}\label{S:Proca}
Proca's equations for a massive classical vector field $A^\mu(x)$ are \cite[p135]{IZ}
\be\label{proca}
\hbar^2\partial _\mu F^{\mu \nu}+m^2 c^2A^\nu=0\,,\quad
F^{\mu \nu}=\partial^\mu A^\nu-\partial^\nu A^\mu\,
\ee
where $m$ is the mass, $\hbar$ is the reduced Planck constant and $c$ is the speed of light. The independent variables $x^\mu$ (for $\mu=0,1,2,3$) are the coordinates of a spacetime event $x$. Following \cite{BjDr64,IZ,Wei95,Zee03}, the metric $g^{\mu\nu}$ for spacetime has the signature $(+ - - -)$\,. One can do no better than to quote \cite[p135]{IZ}, as follows. ``Taking the divergence of this equation we find
\be\label{div}
m^2c^2 \partial \cdot A=0\,.
\ee

``For $m^2\ne 0$, $A$ is divergenceless and (\ref{proca}) reduces to

\be\label{square}
(\hbar^2\square +m^2c^2)A^\mu=0\,,\quad \partial \cdot A =0\,.
\ee

``The vanishing of $\partial \cdot A$ ensures that one of the four degrees of freedom of $A^\mu$ is eliminated in a covariant way, so that the field quanta will indeed carry spin--1". The terse notation in (\ref{div}) is of course $\partial \cdot A= \partial _\mu A^\mu$\,. The well--known relations \cite[p8]{IZ}  between the   tensor $F$ and the electromagnetic fields ${\mathbf E}$ and ${\mathbf B}$ are $F^{0j}=-E^j$ and $F^{jk}=-\epsilon^{jkl}B^l$ where $\epsilon^{jkl}$ is the alternating tensor. 

\subsection{meson algebra}\label{mesalg}

It is now convenient to introduce the scaled fields ${\mathbf e}= {\mathbf E}\hbar/mc$ and ${\mathbf b}= {\mathbf B}/m$\,. Then Proca's equations (\ref{proca}) may be expressed as the Duffin--Kemmer--Petiau equation
\be\label{dkp}
\rmi \hbar\beta_\mu\partial ^\mu \psi+ mc\psi=0
\ee
where $\rmi ^2=-1$\,, and the 10--component \wf is the `DKP spinor' 
\be\label{DKPspin}
 \psi=(A^0,A^1,A^2,A^3,e^1,e^2,e^3,b^1,b^2,b^3) = (A,{\mathbf e},{\mathbf b})\, .
\ee
The notation in (\ref{DKPspin}) for the column vector $\psi$ should include superscripts $t$ denoting transposition, but these are  omitted wherever the meaning is obvious. The four $10 \times 10$ matrices $\beta_\mu$ are compactly expressed as 
\be \label{bet0}
\beta_0= 
\rmi
\left(
\begin{matrix}
0      & \zo^t  & \zo^t & \zo^t\\
\zo   &  Z        & I        &   Z     \\
\zo   & -I         & Z       &   Z     \\
\zo   &  Z        & Z       &   Z     
\end{matrix}
\right)\,,
\ee
 where $\zo=(0,0,0)^t$,  $Z$ is the $3 \times 3$ zero matrix,  $I$ is the $3 \times 3$ unit matrix, while 

\be \label{betj}
\beta_\mathrm{j}= 
\rmi 
\left(
\begin{matrix}
0      & \zo^t  & \ve_\mathrm{j}^{\;t} & \zo^t\\
\zo   &  Z        & Z       &   U_\mathrm{j}    \\
\ve_j   & Z         & Z       &   Z     \\
\zo   &  -U_\mathrm{j}        & Z       &   Z     
\end{matrix}
\right)\,
\ee
 (for $\mathrm{j}=1,2,3$) where $\ve_j$ is the 3-vector with components $(\ve_\mathrm{j})_k=\delta_\mathrm{jk}$\,,  and finally $U_\mathrm{j}$ is the $3 \times 3$ matrix with components $(U_\mathrm{j})_\mathrm{kl}=-\epsilon^\mathrm{jkl}$\,. Notice that $\beta_0=-\beta_0^{\;t}=\beta_0^{\;*t}\equiv \beta_0^{\;\dag}$ is antisymmetric but Hermitian, while $\beta_\mathrm{j}=\beta_\mathrm{j}^{\;t}=-\beta_\mathrm{j}^{\;*t}\equiv -\beta_\mathrm{j}^{\;\dag}$  is symmetric but  antiHermitian. The  matrices $\beta_\mu$ famously \cite{Duff38,Kemm39,Pet36} obey the meson algebra
\be \label{meson}
\beta_\lambda \beta_\mu \beta _\nu + \beta_\nu \beta_\mu \beta _\lambda = \beta_\lambda g_{\mu\nu}+\beta_\nu g_{\mu\lambda}\,.
\ee
As discussed in \cite[p104]{Kemm39}, the three inequivalent and irreducible  representations of (\ref{meson}) have dimension 1, 5 and 10. 

Matrices of particular interest here are
\be \label{bet02}
\beta_0^{\;2}= 
\left(
\begin{matrix}
0      & \zo^t  & \zo^t & \zo^t\\
\zo   &  I        & Z       &   Z     \\
\zo   &  Z       & I        &   Z     \\
\zo   &  Z       & Z       &   Z     
\end{matrix}
\right)\,,
\ee
and
\be \label{eta0}
\eta_0=2\beta_0^{\;2}-1= 
\left(
\begin{matrix}
-1      & \zo^t  & \zo^t & \zo^t\\
\zo    &  I          & Z       &   Z     \\
\zo    & Z         & I         &   Z     \\
\zo    &  Z        & Z        &   -I     
\end{matrix}
\right)\,.
\ee
Moreover,
\be \label{betaeta}
\eta_0\beta_0=\beta_0\eta_0=\beta_0\,, \quad   \eta_0\beta_j+\beta_j\eta_0=0\,,\quad \eta_0\beta_j=\beta_j^{\;\dag}\eta_0\,.
\ee
The first two identities in (\ref{betaeta}) are independent of the representation of the meson algebra (\ref{meson}). 

\subsection{covariance}\label{S:cov}
Covariance of (\ref{dkp}) with respect to the infinitesimal Lorentz transformation ${x^\mu}' \approx x^\mu+\triangle \omega^\mu_{\;\nu}x^\nu$ is established, as in \cite[\S2.2]{BjDr64} for the Dirac equation, by considering $\psi'(x')=S(\triangle\omega)\psi(x)$\, where $S(\triangle\omega)\approx1-(\rmi/4)\triangle\omega^{\mu\nu}r_{\mu\nu}$\,. The generators $r_{\mu\nu}$ satisfy $[\beta_\lambda,r_{\mu\nu}]=2\rmi(g_{\lambda\mu}\beta_\nu-g_{\lambda\nu}\beta_\mu)$\ where $[a,b]$ is the commutator $ab-ba$,  and they are given by $r_{\mu\nu}=2\rmi[\beta_\mu,\beta_\nu]$\,. Defining $\bar{\psi}\equiv\psi^\dag\eta_0$\,, it follows that $\bar{\psi}\phi$ is an invariant bilinear form for any two DKP spinors $\psi$ and $\phi$.

\section{The parametrized Duffin--Kemmer--Petiau wave equation}\label{S:pDKP}
\subsection{one particle}\label{S:one}

\subsubsection{the parametrized DKP equation}\label{S:pdkpe}

The parametrized DKP equation is
\be\label{pdkp}
\beta_\mu\Big[\rmi \hbar\partial ^\mu-(q/c){\mathcal A}^\mu\Big] \psi+ \rmi(\hbar/c)\partial_\tau\psi=0\,.
\ee
Thus the DKP spinor $\psi(x,\tau)$ is now a function of the event $x$ and also the real parameter $\tau$ where $-\infty < \tau < \infty$\,. The minimal substitution\, $\rmi\hbar \partial^\mu \to \rmi \hbar\partial^\mu-(q/c){\mathcal A}^\mu$\, has been assumed, where $q$ is the charge of the particle and ${\mathcal A}^\mu(x)$ is an external potential. The manifestly covariant perturbation series solution of (\ref{pdkp}) is
\be\label{series}
\psi(x,\tau)=\exp[-(\rmi/\hbar) \mathcal{H}(x)(\tau-\tau')]\psi(x,\tau')\,,
\ee
where $\mathcal{H}=-\rmi\hbar c \fy{\partial} +q \fy{\mathcal{A}}$\,. The Feynman slash notation has been appropriated, that is, $\fy{\partial}= \beta \cdot \partial= \beta_\mu \partial ^\mu$\,. 

The discrete symmetries of charge conjugation ${\mathcal C}$\,, parity ${\mathcal P}$ and time reversal ${\mathcal T}$ act  upon the \wf  as follows: 

\be\label{chge}
({\mathcal C}\psi)(x,\tau)= \psi^*(x,-\tau)\,,
\ee
\be\label{part}
({\mathcal P}\psi)(x,\tau)=\eta_0 \psi(x^0, -{\mathbf x},\tau)\,,
\ee
\be\label{time}
({\mathcal T}\psi)(x,\tau)=\eta_0 \psi^*(-x^0, {\mathbf x},\tau)
\ee
and
\be\label{tpc}
(\mathcal {TPC}\psi)(x,\tau)= \psi(-x,-\tau)\,.
\ee
The charge--conjugate \wf  ${\mathcal C}\psi$ satisfies (\ref{pdkp}) subject to the charge  $q$ being replaced with $-q$\,. If a real representation of the matrices $\beta_\mu$ had been chosen then the signs of  $\tau$ in (\ref{chge}) and in (\ref{time}) would be reversed, but would be unaffected in (\ref{part}) and in (\ref{tpc}). The commutation of $\mathcal{TCP}$ conjugation with the vector  interaction in (\ref{pdkp}) extends to the invariant linear interactions of scalar, pseudoscalar,  axial vector and tensor form.

\subsubsection{plane \wfs}\label{S:plane}
 In units where $\hbar=c=1$ and in the absence of an external potential, there are  $\tau$--dependent free plane wave solutions of (\ref{pdkp})  having the forms
\be\label{pwfu}
f^{(+)}_p(x,\tau)=\frac{u(p)}{(2\pi)^2}\exp[\rmi(p \cdot x - \varphi_p m_p\tau)]
\ee
and
\be\label{pwfv}
f^{(-)}_p(x,\tau)=\frac{v(p)}{(2\pi)^2}\exp[\rmi(p \cdot x + \varphi_p m_p\tau)]\,.
\ee

The mass magnitude $m_p$ is the positive square root of $p \cdot p =p_\mu p^\mu$ for subluminal energy-momentum $p$, while $\varphi_p=p^0/|p^0|$  and thus $p^0=\varphi_p E_p$ where $E_p=|p^0|$ is the energy.  The observed rest mass $m_R$ for the particle can be introduced into  a scattering amplitude, for example, through boundary conditions as $\tau \to \pm\infty$\,. The rate of change of coordinate time $x^0$ with respect to the parameter $\tau$ at constant phase $p \cdot x \mp \varphi_p m_p \tau$ is 
\be\label{rate}
\frac{dx^0}{d\tau}=\pm \frac{m_p}{E_p}\,.
\ee
Regardless of the sign of $p^0$\,, for $f^{(+)}_p$ the coordinate time $x^0$ increases  with increasing $\tau$\,, while for   $f^{(-)}_p$ it decreases.

The DKP spinors $u(p)$ and $v(p)$  obey
\be\label{ampequ}
(\fy{p} - \varphi_p m_p)u(p)=0
\ee
and
\be\label{ampeqv}
(\fy{p} + \varphi_p m_p)v(p)=0\,.
\ee
Examination of (\ref{ampequ}) and (\ref{ampeqv}) in the rest frame $p=(p^0,{\mathbf p})=(\varphi_p m_p, \zo)$ shows that there are three degrees of freedom for each mass sign,  consistent with the gauge condition (\ref{div}) and  a representation of spin--1. It is convenient to assemble the three pairs of amplitudes into the $10 \times 3$ matrices 
\be\label{sfu}
{\mathsf u}=
\left(
\begin{matrix}
u^{(1)}& 
u^{(2)} &
u^{(3)}
\end{matrix}
\right)
\ee
and
\be\label{sfv}
{\mathsf v}=
\left(
\begin{matrix}
v^{(1)}& 
v^{(2)} &
v^{(3)}
\end{matrix}
\right)\,,
\ee
given by
\be\label{amppu}
{\mathsf u}(p)=\frac{1}{2 m_p^2}\fy{p}(\fy{p}+\varphi_p m_p){\mathsf w}_+
\ee
and
\be\label{amppv}
{\mathsf v}(p)=\frac{1}{2 m_p^2}\fy{p}(\fy{p}-\varphi_p m_p){\mathsf w}_-\,.
\ee
The  obvious $10 \times 10$ unit matrix $I_{10}$ has been suppressed for clarity. The $10  \times 3$ matrices 
\be\label{wp}
\mathsf{w}_\pm=
\frac{1}{\sqrt{2}}\left(
\begin{matrix}
\zo^t\\
I\\
 \mp \rmi I\\
Z
\end{matrix}
\right)\,
\ee
display the three degrees of freedom for each of $f^{(\pm)}_p$\,. The normalization implies that 
\be\label{norm}
\bar{{\mathsf u}}{\mathsf u}=\bar{{\mathsf v}}{\mathsf v}=I\,,\quad \bar{{\mathsf u}}{\mathsf v}=\bar{{\mathsf v}}{\mathsf u}=Z\,.
\ee 
The associated projections are
\be\label{proju}
\Lambda_u(p)\equiv {\mathsf u}(p) \bar{\mathsf{ u}}(p)=\frac{1}{2m_p^2}\fy{p}(\fy{p}+\varphi_p m_p)
\ee
and
\be\label{projv}
\Lambda_v(p)\equiv {\mathsf v}(p) \bar{\mathsf{ v}}(p)=\frac{1}{2m_p^2}\fy{p}(\fy{p}-\varphi_p m_p)\,.
\ee
The two projections $\Lambda_u(p)$ and $\Lambda_v(p)$ do not sum to  $I_{10}$, thus there is a third orthogonal  projection $\Lambda_z(p)=I_{10}-\Lambda_u(p)-\Lambda_v(p)=I_{10}-\fy{p}\fy{p}/m_p^2$\,. This third projection may be factorized as $\Lambda_z(p)={\mathsf z}(p)\bar{\mathsf{z}}(p)$ where 
\be\label{zamp}
{\mathsf z}(p)=\frac{m^2_p-\fy{p}\fy{p}}{m^2_p}{\mathsf z}_0\,.
\ee
There are four degrees of freedom, expressed as
\be\label{z0}
\mathsf z_0=\left(
\begin{matrix}
1 & \zo^t\\
\zo & Z\\
\zo & Z\\
\zo & I
\end{matrix}
\right)\,,
\ee
such that $\bar{\mathsf z}{\mathsf z } =-I_4$ while $\bar{\mathsf z}{\mathsf u}=\bar{\mathsf z}{\mathsf v}=0$\,, and there are four plane \wfs 
\be\label{f0}
{\mathsf f}^{(0)}_p(x)=\frac{{\mathsf z}(p)}{(2\pi)^2}\exp(\rmi p \cdot x)
\ee
which are independent of $\tau$\,. It follows from (\ref{meson}) that $\fy{p}\fy{p}\fy{p}=m_p^2\,\fy{p}$, hence $\fy{p}\,{\mathsf f}^{(0)}_p(x)=0$ even if $m_p \ne 0$\,. In the rest frame  the nonvanishing components of ${\mathsf f}^{(0)}_p(x)$ consist  of a scalar potential and a magnetic field, both oscillating in coordinate time.

 Any plane wave of the form  $\phi(x,p,\tau)=a \exp[\rmi(p\cdot x +  \varpi \tau)]$\,, where $a$ is any vector in ${\mathbb C}^{10}$\, and $\varpi=\pm\varphi_pm_p$ or $\varpi=0$\,, may be expressed as
\be\label{any}
\phi(x,p,\tau)={\mathsf f}^{(+)}_p(x,\tau){\mathbf c}_+  + {\mathsf f}^{(-)}_p(x,\tau){\mathbf c}_- + {\mathsf f}^{(0)}_p(x){\mathbf c}_0\,,
\ee
where ${\mathbf c}_\pm=\overline{{\mathsf f}^{(\pm)}_p} \phi$\, and ${\mathbf c}_0=\overline{{\mathsf f}^{(0)}_p}\phi $\,.  In particular,  $\phi(x,p,\tau)$ is projected by  $\Lambda_u(p)$  onto the part $\phi^{(+)}(x,p,\tau) \equiv  \Lambda_u(p)\phi(x,p,\tau) ={\mathsf f}^{(+)}(x,p,\tau){\mathbf c}_+$ for which coordinate time $x^0$ increases as the parameter $\tau$ increases\,. 

\subsubsection{influence functions}\label{S:sinfl}
The free \infl $\Gamma^0(x-x',\tau-\tau')$ for (\ref{pdkp}) satisfies
\be\label{infl}
\rmi\fy{\partial}\Gamma^0+\rmi\partial_\tau \Gamma^0=-\delta^4(x-x')\delta(\tau-\tau')\,.
\ee
It follows \cite[p94-95]{Kemm39} that 
\be
\rmi\partial_\tau(\partial_\tau^2-\partial \cdot \partial)\Gamma^0=-(\fy{\partial}\fy{\partial} -\partial \cdot \partial +\partial_\tau^2 -\partial_\tau \fy{\partial})\delta^4(x-x')\delta(\tau-\tau')\,,
\ee
which may be solved using the Fourier transform
\be
\psi(p,\varpi)=\int d^4 x\, d\tau \,\psi(x,\tau)\exp[-\rmi(p\cdot x +\varpi \tau)]
\ee
with inverse 
\be
\psi(x,\tau)=\frac{1}{(2\pi)^5}\int d^4 p\, d\varpi \,\psi(p,\varpi)\exp[+\rmi(p\cdot x +\varpi \tau)]\,.
\ee
The inversion path around the poles in the complex $\varpi$ plane is chosen with, for example,  the mnemonic $(\varpi+\rmi\epsilon)(\varpi^2-m_p^2+\rmi\epsilon)$ leading to 
\begin{multline}\label{infp}
\Gamma^0_+(x-x',\tau-\tau')=\frac{\rmi}{(2\pi)^4}\int d^4p \bigg\{H(\tau-\tau')\frac{\fy{p}(\fy{p}+m_p)}{2m_p^2}\exp[\rmi(p\cdot(x-x')-m_p(\tau-\tau'))] \\-H(\tau'-\tau)\bigg[\frac{\fy{p}(\fy{p}-m_p)}{2m_p^2}\exp[\rmi(p\cdot(x-x')+m_p(\tau-\tau'))]\\+\frac{m_p^2-\fy{p}\fy{p}}{m_p^2}\exp[\rmi p\cdot (x-x')]\bigg] \bigg\}\,,
\end{multline} 
where $H$ is the Heaviside unit step function. For real $p^0$ the `mass' $m_p$ may be real or imaginary.  In the complex plane of $p^0$ there are branch points where $m_p$ vanishes. 

In, for example, the case $\tau-\tau' > 0$ the \infl may be expressed as 
\begin{multline}\label{infL}
\Gamma^0_+(x-x',\tau-\tau')=\frac{\rmi}{(2\pi^4)}\int d^4p\, \bigg\{H(p^0)\Lambda_u(p)\exp[\rmi(p\cdot(x-x')
-\varphi_p m_p(\tau-\tau'))] \\ +H(-p^0)\Lambda_v(p)\exp[\rmi(p\cdot(x-x')+\varphi_p m_p(\tau-\tau'))]\bigg\}
\end{multline}
and as 
\be\label{inff}
\Gamma^0_+(x-x',\tau-\tau')=\rmi \int d^4p\, \bigg\{H(p^0){\mathsf f}^{(+)}_p(x,\tau)\overline{{\mathsf f}^{(+)}_p(x',\tau')}+H(-p^0){\mathsf f}^{(-)}_p(x,\tau)\overline{{\mathsf f}^{(-)}_p(x',\tau')}\bigg\}\,.
\ee

The free evolution of the \wf  $\psi(x,\tau)$   for $\tau >\tau'$ is determined by
\be\label{evop}
\psi_+(x,\tau)=\frac{1}{\rmi}\int d^4x'\,\Gamma^0_+(x-x',\tau-\tau')\psi(x',\tau')\,.
\ee
Positive--energy states project onto the \wfs ${\mathsf f}^{(+)}_p$ for all of which coordinate time $x^0$ increases with $\tau$\,, while negative--energy states project onto  the \wfs ${\mathsf f}^{(-)}_p$ for all of which $x^0$ decreases. For energy of either sign, the frequency with respect to $\tau$ is $\varpi= -m_p$ which is by convention `positive mass'. The alternative mnemonic $(\varpi+\rmi\epsilon)(\varpi^2-m_p^2-\rmi\epsilon)$ leads to  the `negative mass' $\varpi =+ m_p$\,. Note that the subscript $+$ in (\ref{evop}) indicates the choice of inversion path and thus the mass positivity, not the sense of change of coordinate time.

The advanced influence function $H(\tau'-\tau)\Gamma^0_+(x-x',\tau-\tau')$  has simple poles in the complex plane of $p^0$\,, at $p^0=\pm|{\mathbf p}|$ where $m_p=0$\,.  These massless particles  satisfy the DKP equation (\ref{dkp}) with $m=0$\,. They are not photons since they do not satisfy Maxwell's equations. For a 10--dimensional DKP--like formulation of the latter equations, see \cite{HarCh46,Gho01}\,.

The \infl $\Gamma^0_+$ may also be constructed using
\be\label{GD}
\rmi \partial _\tau \Gamma^0_+(x-x',\tau-\tau')=-(\fy{\partial}\fy{\partial} -\partial \cdot \partial +\partial_\tau^2 -\partial_\tau \fy{\partial})\Delta^0_+(x-x',\tau-\tau')
\ee
where $\Delta^0_+$ is the \infl for the wave equation with four spacelike dimensions, that is,
\be\label{Del}
\left(\frac{\partial ^2}{\partial t^2}-\frac{\partial^2}{\partial \tau^2}-\nabla^2\right)\Delta^0_+(x-x',\tau-\tau')=-\delta^4(x-x')\delta(\tau-\tau')
\ee
with $t=x^0$ being the timelike variable. There are many closed forms for $\Delta^0_+$\,, which have all been shown \cite{AhHo2011} to be equivalent to
\be\label{AhHo}
\Delta^0_+(t,{\mathbf x},\tau)=\frac{1}{4\pi^2 r}\frac{\partial}{\partial r}\frac{H(t^2-\tau^2-r^2)}{\sqrt{t^2-\tau^2-r^2}}
\ee 
where $r^2={\mathbf x}\cdot{\mathbf x}$\,. The influence of $\Delta^0_+(t,{\mathbf x},\tau)$ and hence also that of $\Gamma^0_+(t,{\mathbf x},\tau)$ is restricted to the surface and interior of the light cone at $(t,{\mathbf x})=(0,{\mathbf 0})$\,.

The \infl  in the presence of an external potential ${\mathcal A}^\mu(x)$ is $\Gamma(x,\tau;x',\tau')$\,, which satisfies 
\be\label{inflA}
(\rmi\fy{\partial}-q\fy{\mathcal A})\Gamma+\rmi\partial_\tau \Gamma=-\delta^4(x-x')\delta(\tau-\tau')\,.
\ee
As demonstrated in \cite{Benn2014}, a solution  $\Gamma_+$ may be constructed by iteration on the free \infl $\Gamma^0_+$\,.

\subsection{two particles}\label{S:many}
The parametrized DKP equation (\ref{dkp}) is extendable to many particles. For two independent particles at events $x$ and $y$ the \wf $\Psi(x,y,\tau)$  is a linear combination of  single--particle \wf products such as $\psi_1(x,\tau)\otimes \psi_2(y,\tau)$\,, and satisfies
\be\label{two}
(\rmi \fy{\partial }_x-q_1\fy{\mathcal A}(x)) \otimes I_{10}\Psi(x,y,\tau)+ I_{10}\otimes  (\rmi \fy{\partial }_y-q_2\fy{\mathcal A}(y)) \Psi(x,y,\tau)+\rmi\partial_\tau\Psi(x,y,\tau)=0\,.
\ee

If the particles are identical then $\Psi$ is either antisymmetric with respect to the exchange of \wfs: 
\be\label{asymm}
\Psi(x,y,\tau)={\mathcal N}\big\{\psi_1(x,\tau)\otimes \psi_2(y,\tau)-\psi_2(x,\tau)\otimes\psi_1(y,\tau)\big\}\,,
\ee
or symmetric: 
\be\label{symm}
\Psi(x,y,\tau)={\mathcal N}\big\{\psi_1(x,\tau)\otimes \psi_2(y,\tau)+\psi_2(x,\tau)\otimes \psi_1(y,\tau)\big\}\,,
\ee
where ${\mathcal N}$ denotes some normalizing constant. The two--particle influence function in the presence of an external potential is chosen to be
\be\label{twoinfl}
\Gamma_{++}(x,y,\tau;x',y',\tau')=\frac{1}{\rmi}\Gamma_+(x,\tau;x',\tau')\otimes \Gamma_+(y,\tau;y',\tau')\,.
\ee
The first factor on the right hand side of (\ref{twoinfl}) depends upon the charge $q_1$, the second upon $q_2$\,. In the case of identical particles, evolution under the influence of $\Gamma_{++}$ preserves antisymmetry (\ref{asymm}) and symmetry (\ref{symm}).

The conservation of current consistent with (\ref{two}) is (assuming again that $\hbar=c=1$)
\be \label{concur}
\frac{\partial}{\partial x_\mu} j_\mu^x + \frac{\partial}{\partial y_\mu} j_\mu^y +\frac{\partial}{\partial \tau} \overline{\Psi}\Psi=0\,,
\ee 
where $\overline{\Psi}=\Psi ^\dg\eta_0 \otimes \eta_0$\,. The currents are
\be\label{curx}
j^x_\mu(x,y,\tau)=\overline{\Psi}(x,y,\tau)\beta_\mu \otimes I_{10}\Psi(x,y,\tau)
\ee
and
\be\label{cury}
j^y_\mu(x,y,\tau)=\overline{\Psi}(x,y,\tau)I_{10} \otimes \beta_\mu \Psi(x,y,\tau)\,.
\ee

The parametrized formulation may be used to calculate scattering amplitudes and bound states, just as for the parametrized Dirac wave equation \cite{Benn2014}. Standard results for Dirac particles obtain \cite{Benn2014} if the currents (\ref{curx}) and (\ref{cury}) are concatenated, that is, the currents are integrated over all $\tau$ for the purpose of calculating electromagnetic fields of interaction \cite{ArHoLa83}.

\section{Preparation}\label{S:prep}

\subsection{eigenstates of rotation}\label{S:rot}
For any pure spacelike 4-vector $s=(0,{\mathbf s})$ of unit length, rotation about that vector through an angle $\theta$ induces the operation $O({\mathbf s},\theta)$ on DKP spinors where  
\be\label{rotop}
O({\mathbf s},\theta)=
\left(
\begin{matrix}
1 & \zo^t & \zo^t &\zo^t\\
\zo & \exp(\rmi \theta {\mathbf s} \cdot {\mathbf V})& Z &Z\\
\zo & Z &  \exp(\rmi \theta {\mathbf s} \cdot {\mathbf V}) & Z\\
\zo & Z & Z & \exp(\rmi \theta {\mathbf s} \cdot {\mathbf V}) 
\end{matrix}
\right)
\ee
and $V_\mathrm{j}=\rmi U_\mathrm{j}$\,. The eigenvectors ${\mathbf v}_\mathrm{l}$   of ${\mathbf s} \cdot {\mathbf V}$   satisfy $\rmi {\mathbf s}\times{\mathbf v}_\mathrm{l}=\mathrm{l}{\mathbf v}_\mathrm{l}$\,, where the eigenvalues are $\mathrm{l}=0$\,, $-1$ and $1$\,. It is readily seen that the eigenvectors are orthogonal  and may be normalized, such that ${\mathbf v}_\mathrm{k}^*\cdot {\mathbf v}_\mathrm{l}=\delta_\mathrm{kl}$ where $\mathrm{k}$ and $\mathrm{l}=0,-1,1$\,. The eigenvalues of $O$ are 1, $\exp(-\rmi \theta)$\, and $\exp(+\rmi \theta)$\,.  The eigenvectors of $O$ provide an orthonormal 10--spinor basis in the form of the columns of the matrix $X =(\chi_1,\chi_2,\chi_3,\chi_4,\chi_5,\chi_6,\chi_7,\chi_8,\chi_9,\chi_{10})$ where
\be\label{basis}
X=
\left(
\begin{matrix}
1&0&0&0&0&0&0&0&0&0\\
\zo&{\mathbf v}_0&\zo&\zo&{\mathbf v}_{-1}&\zo&\zo&{\mathbf v}_{+1}&\zo&\zo\\
\zo&\zo&{\mathbf v}_0&\zo&\zo&{\mathbf v}_{-1}&\zo&\zo&{\mathbf v}_{+1}&\zo\\
\zo&\zo&\zo&{\mathbf v}_0&\zo&\zo&{\mathbf v}_{-1}&\zo&\zo&{\mathbf v}_{+1}\\
\end{matrix}
\right)\,.
\ee
For example, $\chi_5=(0,{\mathbf v}_{-1}^{\;\;\;\;t},\zo^t,\zo^t)^t$\,. Any 10--spinor $\zeta$ may be expanded as
\be\label{expz}
\zeta=\sum_{d=1}^{10}K_d\chi_d 
\ee
where the scalar coefficients are $K_d=\chi_d^\dg \zeta$\,.

Owing to the completeness of the basis $X$ in ${\mathbb C}^{10}$\,, it may be assumed  for the purpose of this investigation that a single particle with \wf $\zeta(x,\tau)$ was prepared at the parameter value  $\tau =0$ in an eigenstate of the rotation operator $O({\mathbf s},\theta)$\,. That is, $\zeta(x,0)=\zeta_0(x)=K_d(x)\chi_d$ for some $d$, where $O\chi_d=\exp(\rmi\, l_d\, \theta)\chi_d$ and $l_d=0,-1$ or $+1$\,. In general $O$ does not commute with  the influence funtion $\Gamma_+$\,, and so $\zeta(x,\tau)$ is not in a rotational eigenstate for $\tau>0$\,. 
\

\subsection{identical particles}\label{S:twoid}
Consider now two particles prepared at $\tau=0$ in eigenstates of rotation $O({\mathbf s},\theta)$ with respect to the same pure spacelike direction $s=(0,{\mathbf s})$\,. The eigenfunctions  are denoted $\zeta_0(x)$ and $\phi_0(x)$\,. The eigenvalues  $\exp(\rmi  k   \theta)$ and $\exp(\rmi l \theta)$\, of the two eigenstates need not be the same. As discussed at length by Jabs \cite{Jabs10,Jabs2014}, it is not necessarily the case that the common direction of rotation is in a common rotational frame, and thus the DKP--spinor amplitudes  of the two eigenfunctions  must include distinct factors $\exp(\rmi k \kappa)$ and $\exp(\rmi l \xi)$\,, where $\kappa$ and $\xi$ are angles in the interval $[0,2\pi]$ on a plane perpendicular to ${\mathbf s}$\,,  and the angles are relative to some pure spacelike reference direction in the plane. The frame--dependent phase factors may alternatively be displayed explicitly by expressing the eigenstates as $\exp(\rmi k \kappa)\zeta_0(x)$ and $\exp(\rmi l \xi)\phi_0(x)$\,. Thus, in the case of identical particles at the events $x$ and $y$\,, a  preliminary two--particle \wf  $\hat{\Psi}_0$ is obtained as a  sum  of products of single--particle \wfs with \wfs exchanged, except for the angles $\kappa$ and $\xi$\,, as in 
\begin{multline}\label{prelim}
\hat{\Psi}_0(x,y,\kappa,\xi)=\big\{\exp(\rmi k \kappa)\zeta_0(x)\otimes \exp(\rmi l \xi)\phi_0(y) \\ +\exp(\rmi l \kappa)\phi_0(x)\otimes \exp(\rmi k \xi)\zeta_0(y)\big\}\,.
\end{multline}
A normalization factor has been omitted. The prepared eigenfunction $\zeta_0(x)$ evolves under the influence of $\Gamma_+$ into some \wf $\zeta(x,\tau)$ for $\tau > 0$\,, and $\phi_0(y)$ evolves into some $\phi(y,\tau)$\,. Assume that, say,  $\kappa>\xi$\,. Change $\xi$ to $\kappa$ in the second summand in (\ref{prelim}) by an anticlockwise rotation through $\kappa-\xi$\,. Next, also in the second summand, change the original $\kappa$ to $\xi$  in an homotopically consistent fashion \cite{Jabs10,Jabs2014}.  That is, make the change by  anticlockwise rotation through $2\pi-\kappa+\xi$. After particle exchange the final two--particle \wf  becomes
\begin{multline}\label{final}
\Psi_0(x,y,\mu,\lambda)=\big\{\exp(\rmi k \kappa)\zeta_0(x)\otimes \exp(\rmi l \xi)\phi_0(y) \\ +\exp(2 l \pi \rmi)\exp(\rmi l \xi)\phi_0(x)\otimes \exp(\rmi k \kappa)\zeta_0(y)\big\}\,.
\end{multline}
The factors $\exp(\rmi k \kappa)$ and $\exp(\rmi l \xi)$ may be reabsorbed into the \wfs $\zeta_0$ and $\phi_0$ respectively, and have no further physical significance here.  The factor of $\exp(2 l \pi \rmi)$ in the second term does have physical significance. Again $l=-1,0,1$\,, and so the \wf is  symmetric under particle exchange. Moreover,  $\Psi_0(x,y,\mu,\lambda)$ evolves for $\tau>0$ into
\be\label{later}
\Psi(x,y,\tau,\mu,\lambda)=\big\{\zeta(x,\tau)\otimes \phi(y,\tau) +\exp(2 l \pi \rmi)\,\phi(x,\tau)\otimes\zeta(y,\tau)\big\}
\ee
which retains the symmetry. 

The extensions of (\ref{two}) and (\ref{twoinfl}) to $N > 2$  particles is obvious. The extension of (\ref{final}) to $N>2$ particles  is straightforward and may be found in 
\cite{Jabs2014}\,. It follows \cite[p129-130]{WeiQM13} that for any value of the parameter $\tau>0$ a grand canonical ensemble of spin--1 particles has Bose--Einstein statistics.

\section{Entanglement}\label{S:ent}
If the two--particle \wf were a simple product of normalized single--particle wavefunctions, such as 
\be\label{simple}
\Psi(x,y,\tau)=\zeta(x,\tau)\otimes\phi(y,\tau)\,,
\ee
then the current in the  spacetime of  the first particle  would be
\be\label{currsimpone}
j^x_\mu(x,y,\tau)= \overline{\zeta}(x,\tau)\beta_\mu\zeta(x,\tau)\;\overline{\phi}(y,\tau)\phi(y,\tau)\,.
\ee
Integrating over the spacetime of the second particle would yield the current of the first particle as though it were alone, that is
\be\label{intone}
j^x_\mu(x,\tau)=\int j^x_\mu(x,y,\tau)\,d^{\,4}y= \overline{\zeta}(x,\tau)\beta_\mu\zeta(x,\tau)\,.
\ee
The marginal current for  the second particle would be
\be\label{currsimptwo}
j^y_\mu(y,\tau)=\int j^y_\mu(x,y,\tau)\,d^{\,4}x= \overline{\phi}(y,\tau)\beta_\mu\phi(y,\tau)\,.
\ee

The two--particle state with \wf (\ref{later}) is entangled in the sense that $j^x_\mu$  and $j^y_\mu$ are  not simple products. Rather the currents  are (dropping the subscript $+$\,, suppressing the normalization ${\mathcal N}$ and specifying  that the eigenvalue $l$ is $0$ or $\pm1$)
\begin{multline}\label{entcur}
j^x_\mu(x,y,\tau)= j^y_\mu(y,x,\tau)=\overline{\zeta}(x,\tau)\beta_\mu\zeta(x,\tau)\;\overline{\phi}(y,\tau)\phi(y,\tau)\\+ \overline{\zeta}(x,\tau)\beta_\mu\phi(x,\tau)\;\overline{\phi}(y,\tau)\zeta(y,\tau)\\+ \overline{\phi}(x,\tau)\beta_\mu\zeta(x,\tau)\;\overline{\zeta}(y,\tau)\phi(y,\tau)\\+ \overline{\phi}(x,\tau)\beta_\mu\phi(x,\tau)\;\overline{\zeta}(y,\tau)\zeta(y,\tau)\,.
\end{multline}
The \wfs $\zeta$ and $\phi$ for two identical particles have the same intrinsic parameters such as mass and charge, but may have different spin polarizations or different centers  \cite{Jabs10,Jabs2014} and so may respectively be orthogonal or have no overlap. Entanglement does not vanish in the absence of projection or overlap, since each marginal current remains influenced by the existence of the other particle. That is, neglecting the $y$--integrals of the second and third summands in (\ref{entcur}),   
\be\label{currs}
j^x_\mu(x,\tau)=j^y_\mu(x,\tau) \approx \overline{\zeta}(x,\tau)\beta_\mu\zeta(x,\tau)+\overline{\phi}(x,\tau)\beta_\mu\phi(x,\tau)\,.
\ee

To establish the connection between the proof that DKP particles are bosons and the preservation of unrestricted entanglement  as in (\ref{later}), consider  the  total value of  the energy--momentum of the first particle defined as 
\be\label{expp}
\langle p^\mu \otimes I_{10}\rangle = \int \int \,\overline{\Psi}\, p^\mu \otimes I_{10}\,\Psi\,d^{\,4}x \, d^{\,4}y\, \,.
\ee
The value must be independent of $\tau$ for free particles if  energy--momentum is to be conserved. The required independence  follows from  the tensor structure of  (\ref{two}). Consider next relativistic angular momentum. For the first of two particles the angular momentum operator is $M_{\mu\nu}\otimes I_{10}$\,, where 
\be\label{angmtm}
M_{\mu\nu}=x_\mu p_\nu-x_\nu p_\mu +\frac{1}{\rmi} (\beta_\mu\beta_\nu-\beta_\nu\beta_\mu)
\ee
is the sum of orbital and spin components \cite{Duff38}.  The  conservation of total angular momentum is also a consequence of the tensor structure of (\ref{two}). The conservation of concatenated currents, that is, the integral of (\ref{concur}) over all $\tau$, implies that the concatenated marginal currents can be the sources for electromagnetic fields.  Again, (\ref{concur}) is a consequence of the tensor structure of (\ref{two}). It may be conjectured that no other structure for (\ref{two}) yields these essential conservation laws. The preservation of unrestricted entanglement  is a further consequence of that structure.

\section{Discussion}\label{S:disc}
 The proof of the spin--statistics connection according to Quantum Field Theory requires that fields of creation and destruction operators  for particles of integer (half--odd--integer) spin   commute (anticommute) if at spacelike separations  \cite{Wei95}. These commutation and anticommutation relations for fields further ensure the covariance of coordinate--time--ordered  Dyson series \cite{Wei95} and $\mathcal{TPC}$ invariance \cite{Sozzi08}\,.  However, owing to the causality conditions upon  the field commutators and anticommmutators,  the quantum vacuum  is the most entangled and even that entanglement is restricted to  separations of a few Compton wavelengths $\hbar/mc$ or Compton periods $\hbar/mc^2$ \cite{Haag58,Araki62,Fred85,Summ11,Ols12}\,.

There are no causality conditions upon the commutators for the first--quantized operators such as  spacetime position $x^\mu$   and energy--momentum $p^\mu$\,,  yet the parametrized DKP wave equation yields the spin--statistics connection for particles of spin--0 and spin--1.   The covariance of series solutions for the equation and the invariance  of solutions under $\mathcal{TPC}$ transformation are both manifest, and the formulation allows unrestricted entanglement.

\appendix

\setcounter{section}{1}

\section*{Appendix}\label{S:app}

\subsection{spin--0}\label{S:zero}

The DKP five-spinor for a spin--0 particle is $\psi=(\partial^0\rho,\partial ^1\rho,\partial^2\rho, \partial^3\rho,-m\rho$)\,, where $\rho$ satisifies the Klein--Gordon equation \cite[p5]{BjDr64}
\be\label{KG}
\partial_\mu\partial^\mu\rho+m^2\rho=0\,.
\ee
 The four $5 \times 5$ DKP matrices are 
\be \label{0bet0}
\beta_0= 
\rmi
\left(
\begin{matrix}
0      & \zo^t  & -1  \\
\zo   &  Z      & \zo \\
1      & \zo^t &    0  
\end{matrix}
\right)
\ee
and (for $\mathrm{j}=1,2,3)$ 
\be \label{0betj}
\beta_\mathrm{j}= 
\rmi 
\left(
\begin{matrix}
0      & \zo^t  & 0          \\
\zo   &  Z       & \ve_\mathrm{j}    \\
0   & \ve_\mathrm{j}^{\;t}  &   0        
\end{matrix}
\right)\,.
\ee
These obey the meson algebra (\ref{meson}), and also (\ref{betaeta}) where now
\be \label{0eta0}
\eta_0=2\beta_0^{\;2}-1= 
\left(
\begin{matrix}
1      & \zo^t  & 0\\
\zo    &  -I      & \zo  \\
0    &  \zo^t        &1     
\end{matrix}
\right)\,.
\ee
The single eigenstates of the projections $\Lambda_u$ and $\Lambda_v$ are 
\be\label{0wp}
\mathsf{w}_\pm=
\frac{1}{\sqrt{2}}\left(
\begin{matrix}
1\\
\zo\\
\pm \rmi
\end{matrix}
\right)\,,
\ee
in the rest frame, while the three eigenstates of $\Lambda_z$ are 
\be\label{0zp}
\mathsf{z}_0=\left(
\begin{matrix}
\mathbf{0}^t\\
I\\
\mathbf{0}^t
\end{matrix}
\right)\,,
\ee
also in the rest frame.

 The rotation operator  $O({\mathbf s}, \theta)$ is now
\be\label{0rot}
O({\mathbf s},\theta)=
\left(
\begin{matrix}
1 & \zo^t & 0\\
\zo & \exp(\rmi {\mathbf s} \cdot {\mathbf V}) &\zo \\
0 & \zo^t & 1
\end{matrix}
\right)\,.
\ee
As pointed out by Duffin \cite{Duff38}, the spatial gradient  of $\rho$ responds to a rotation. The five eigenvectors of $O$ are 
\be\label{five}
X=
\left(
\begin{matrix}
1 &0&0&0&0\\
0&{\mathbf v}_0&0&{\mathbf v}_{-1}&{\mathbf v}_{+1}\\
0&0&1&0&0\\
\end{matrix}
\right)\,.
\ee

\subsection{ spin--1/2}\label{S:half}

The influence function $\Gamma_+$ for the parametrized Dirac wave equation may be found in \cite{Benn2014,Benn2015}. The influence function for two particles, also given by (\ref{twoinfl}), preserves  any symmetry or antiysmmetry of the two--particle wavefunction.
The rotation operator is, in terms of the Dirac matrices, 
\be\label{rothalf}
O({\mathbf s},\theta)=\exp(\rmi\gamma^0\fy{s}\gamma^5/2)\,.
\ee
Its two eigenvalues $\exp(\rmi l \theta)$, where $l=\pm1/2$,  are doubly--degenerate. The projection operators for $l=\pm1/2$ are
\be\label{proj}
P({\mathbf s},l)=1/2 + l \gamma^0\fy{\mathbf s}\gamma^5\,.
\ee
That is, $P({\mathbf s},l)$ projects any 4-spinor $\psi$ onto a rotational eigenstate:
\be
O(s,\xi)=\exp\left(\frac{\rmi}{2} \xi \gamma^0 \gamma^5 \fy{s}\right)\,.
\ee
The \wf at $\tau=0$ for two identical particles is also given by (\ref{later}). The factor before the second summand in (\ref{later}) is now $\exp( 2 \pi \rmi l)=-1$ and so Fermi--Dirac statistics are inferred.

\section*{References}

\bibliography{DuffKemmPet.bbl}

\end{document}